\input{aipcheck}

\documentclass[
    ,final            
  ]
  {aipproc}

\layoutstyle{6x9}

\begin{document}

\title{The GRB-SN connection: GRB~030329 and XRF~030723}

\author{J. P. U. Fynbo}{
  address={Department of Physics and Astronomy, Ny Munkegade,
  DK-8000 Aarhus C, Denmark}
}
\author{J. Hjorth}{
  address={Astronomical Observatory, 
  Juliane Maries Vej 30, DK--2100 Copenhagen~\O, Denmark}
}
\author{J. Sollerman}{
  address={Stockholm Observatory, Department of Astronomy, AlbaNova,
  S-106 91 Stockholm, Sweden}
}
\author{P. M\o ller}{
  address={European Southern Observatory, Karl Schwarzschild-Strasse 2,
  D-85748 Garching, Germany}
}
\author{J. Gorosabel}{
  address={Instituto de Astrof\'\i sica de Andaluc\'\i a, IAA-CSIC,
  P.O. Box 03004, 18080 Granada, Spain}
}
\author{F. Grundahl}{
  address={Department of Physics and Astronomy, Ny Munkegade,
  DK-8000 Aarhus C, Denmark}
}
\author{B.~L.~Jensen}{
  address={Astronomical Observatory,
  Juliane Maries Vej 30, DK--2100 Copenhagen~\O, Denmark}
}
\author{Michael I. Andersen}{
  address={Astrophysikalisches Institut Potsdam, An der Sternwarte 16,
  D-14482 Potsdam, Germany}
}
\author{P. Vreeswijk}{
  address={European Southern Observatory, Casilla 19001, Santiago 19,
  Chile}
}
\author{A.~Castro-Tirado}{
  address={Instituto de Astrof\'\i sica de Andaluc\'\i a, IAA-CSIC,
   P.O. Box 03004, 18080 Granada, Spain}
}
\author{the GRACE collaboration}{
address={http://zon.wins.uva.nl/~grb/grace/}
}

\begin{abstract}
The attempt to secure conclusive, spectroscopic evidence for the 
GRB/SN connection has been a central theme in most GRB observing time 
proposals since the discovery of the very unusual GRB~980425 associated
with the peculiar type Ib/c SN~1998bw. GRB~030329 provided this evidence 
to everybody's satisfaction. In this contribution we show the results of 
a spectroscopic campaign of the supernova associated with GRB~030329 
carried out at ESOs Very Large Telescope. We also present preliminary 
results from a photometric and spectroscopic campaign targeting the 
X-ray Flash of July 23.
\end{abstract}

\maketitle


\section{Introduction}
 
Like GRB~980425, the Gamma Ray Burst (GRB) detected by the Hete-II satellite
on March 29, 2003 was born famous. It was so bright in $\gamma$-rays that the 
duty astronomer on the Hete-2 team designated it 'monster GRB'. The fluence 
alone places GRB~030329 in the top 0.2\% of the 2704 GRBs detected with 
the Burst And Transient Source Experiment (BATSE) during its nine years of 
operation. The designation was further justified by the detection of the very 
bright Optical Afterglow (OA) with an optical magnitude around 12.4 (e.g. Price 
et al. 2003; Torii et al. 2003; Sato et al. 2003). Its redshift was determined
with the high resolution UVES spectrograph at the European Southern Observatory
(ESO) to be $z=0.1685$ (Greiner
et al. 2003a), the lowest redshift ever measured for a normal (excluding 
GRB~980425), long duration GRB. From that point on it was clear to everybody 
that GRB~030329 offered a unique chance to finally obtain 
spectroscopic proof of the connection between long duration GRBs and core
collapse supernovae (SNe). This connection was believed to exist both from
theoretical expectation (e.g. MacFadyen \& Woosley 1999; Dado et al. 2003 and 
references therein) and observational hints. The strongest but also most elusive 
observational hint came from the very unusual GRB~980425 that was associated
with the very energetic type Ib/c SN~1998bw (Galama et al. 1998; Patat et al.
2001). However, the fact that GRB~980425 had a total equivalent isotropic energy 
release four orders of magnitude smaller than any other well studied long
duration GRB left the reasonable doubt that it possibly was not representative
for other GRBs.

Another important set of observations were bumps seen superimposed on the
OA lightcurves (e.g. Bloom et al. 1999; Castro-Tirado \&
Gorosabel 1999). For GRB~011121 and GRB~021211 there were, in 
addition to photometric evidence, also tentative but not conclusive 
spectroscopic evidence for an underlying core collapse SN
(Garnavich et al. 2003a; Greiner et al. 2003b; Della Vella et al. 2003).
The attempt to secure conclusive, spectroscopic evidence for the GRB/SN
connection was therefore a central theme in GRB science,
especially from 1999 and onwards. 

In this contribution we describe the observations of the afterglow and 
SN associated with GRB~030329 within the context of 
GRACE\footnote{Gamma Ray Afterglow Collaboration at ESO, 
http://zon.wins.uva.nl/~grb/grace/.}.
These results have been published in Hjorth et al. (2003) and will
be more thoroughly described in a paper in preparation. We 
also briefly review the results obtained by other groups. Finally, we show 
preliminary results from an extensive GRACE follow-up of the X-Ray Flash 
(XRF) from July 23 2003. In this case there is photometric 
evidence for an associated SN.

\section{GRACE observations of OA~030329}
After the determination of the redshift of GRB~030329, we designed an
extensive spectroscopic campaign aimed at detecting and following the
supernova expected to be associated with the GRB. Spectra were obtained
with the FORS1 or FORS2 spectrographs on six epochs from April 3 through 
May 1. In the left panel of Fig.~1 we show the six flux-calibrated 
spectra. As seen, the spectrum evolves from a featureless power-law 
spectrum to a SN-like spectrum dominated by broad features. The presence
of a SN was first reported by Garnavich et al. (2003b) on April 9 2003 whereby 
it was designated SN~2003dh. Superimposed
on the spectrum are several strong emission lines from the underlying
host galaxy (shown in more detail in the right panel of Fig.~1).
Shown in the left panel of Fig.~1 with a dashed line
is the spectrum of SN~1998bw 33 days after GRB~980425. The similarity 
between this spectrum and the spectrum of SN~2003dh is striking.

In the left panel of Fig.~2 we show the V-band lightcurve of 
OA~030329 primarily based on observations from ESO telescopes by
GRACE (Guziy et al. 2004, submitted). There is no obvious
SN bump seen in the lightcurve. 

\begin{figure*}
  \includegraphics[height=.28\textheight]{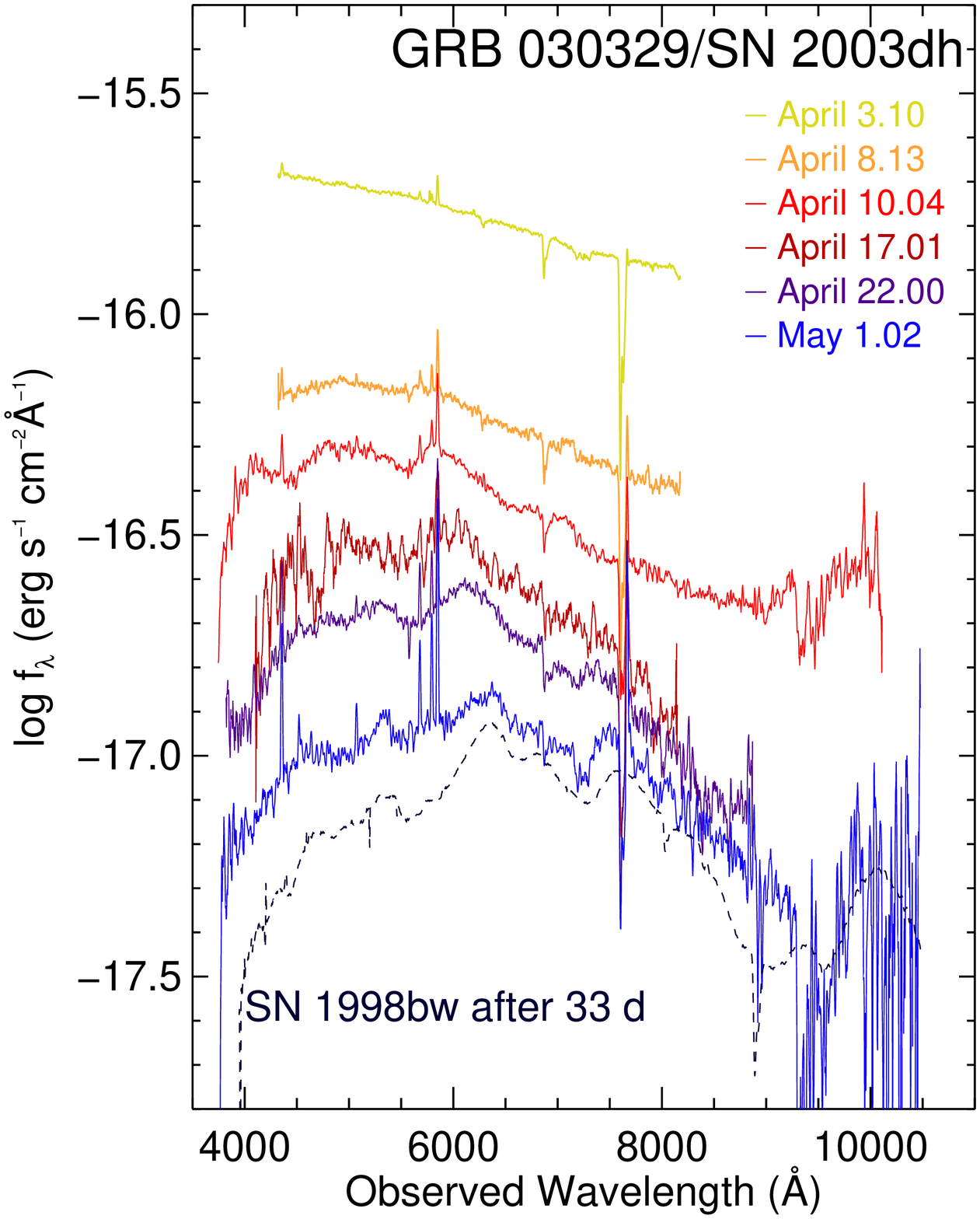}
  \includegraphics[height=.28\textheight]{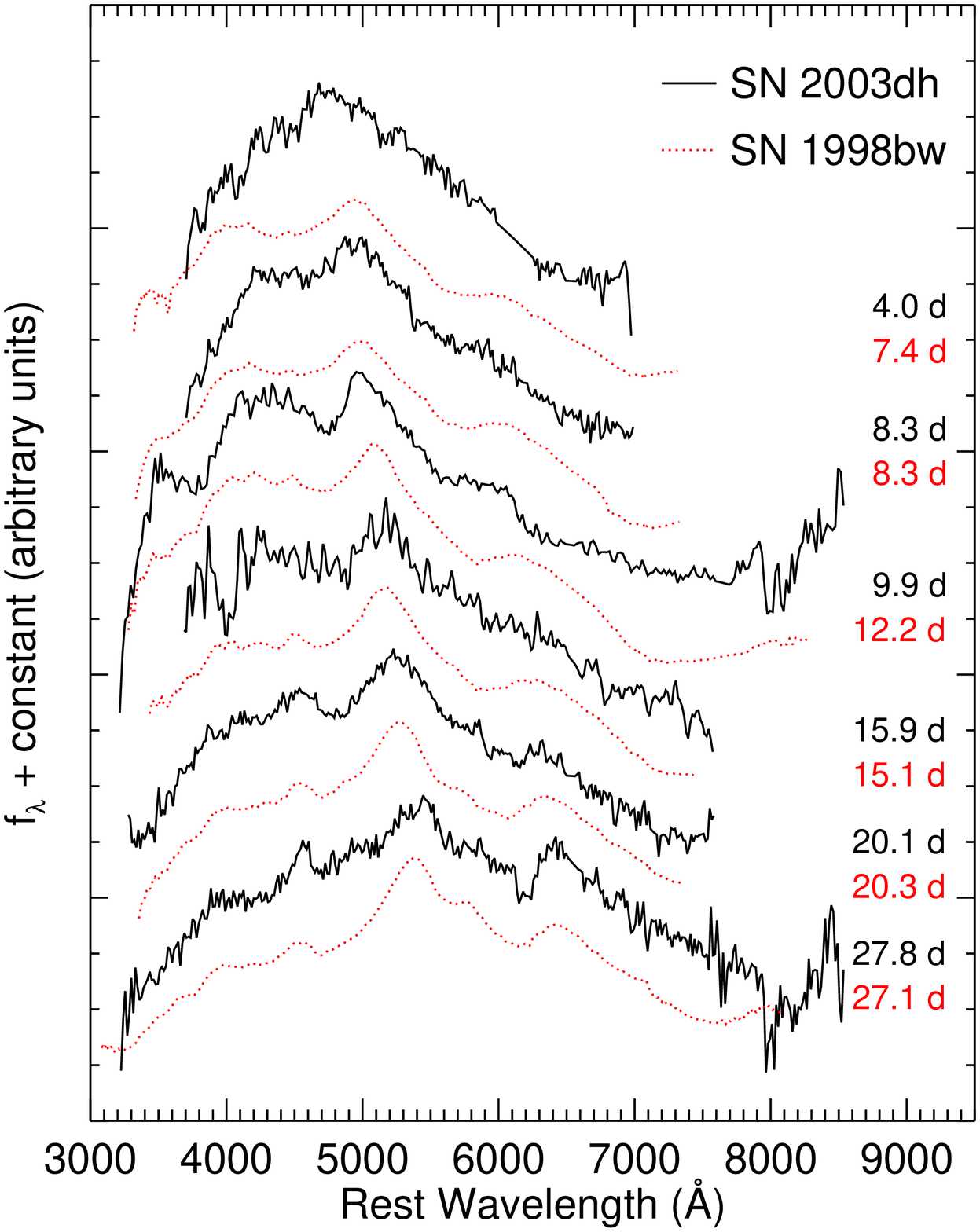}
  \includegraphics[height=.28\textheight]{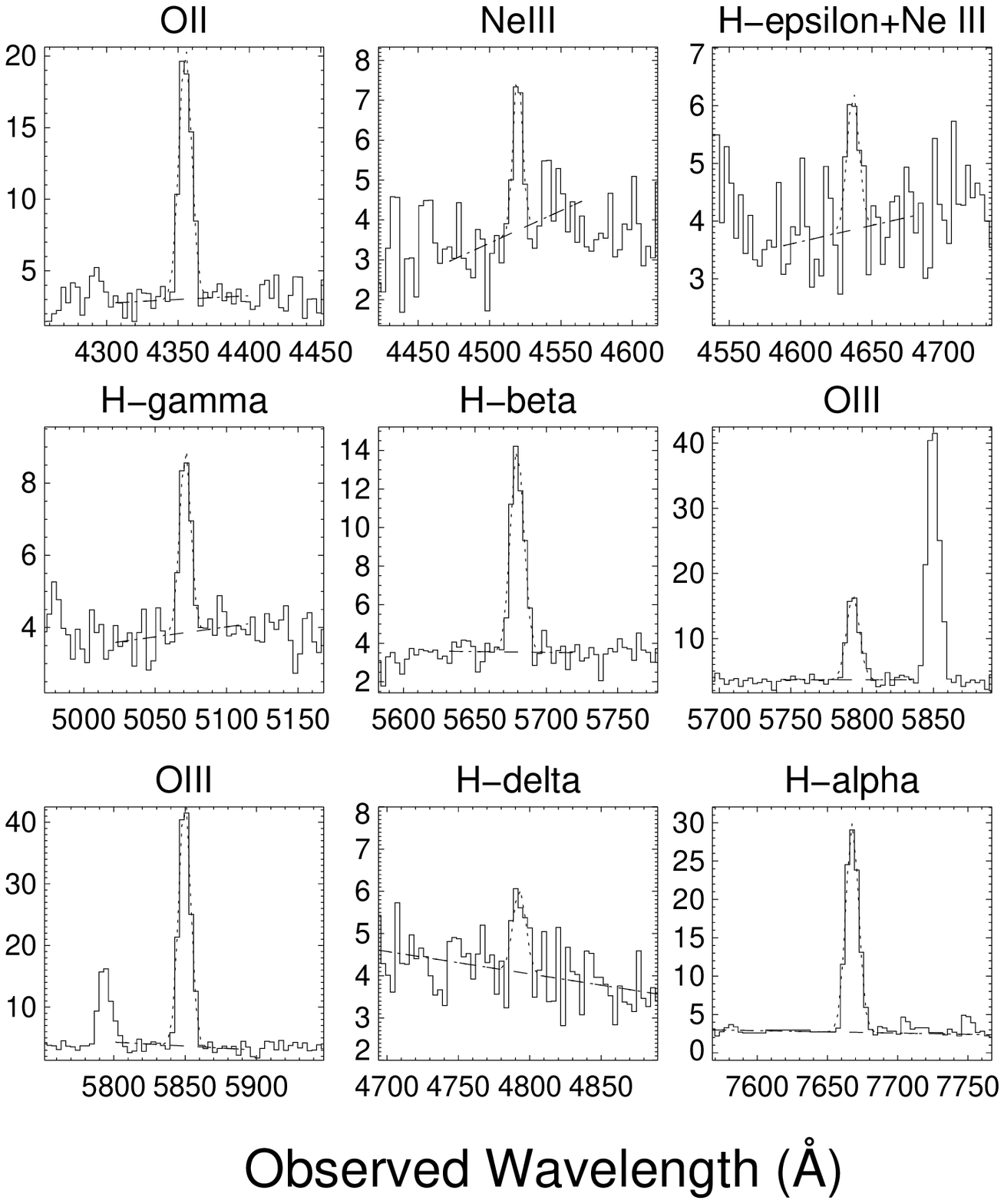}
  \caption{}
\end{figure*}

To investigate the SN component in more detail we performed a spectral 
decomposition of the afterglow, SN and host galaxy components
as follows. While the host galaxy has strong emission lines, its continuum 
flux upper limit is negligible at early epochs and significantly less than 
the total flux at the later epochs. The contribution from the host galaxy was 
therefore accounted for by simply removing the emission lines. Model spectra 
were constructed as a sum of a power law (f$^{-\beta}$) and a scaled version 
of one of the SN~1998bw template spectra from Patat et al.\ (2001). For each 
template, or section thereof, a least-squares fit was obtained through fitting 
of the three parameters: power-law index $\beta$, amplitude of afterglow, and 
amplitude of supernova. In most cases the best fitting index was found to be
$\beta=-1.2\pm0.05$ which was adopted throughout. We note, however, that the 
resulting overall spectral shape of the supernova contribution does not depend 
on the adopted power-law index or template spectrum. The result of this 
decomposition is shown in the middle panel of Fig.~1. The striking similarity 
between the spectra of SN~1998bw and SN~2003dh is clearly seen. The spectral peak 
wavelength, for both supernovae, is shifting towards the red. The shift is on 
average 25\AA \  per day for SN~2003dh, which is similar to the evolution of 
the early spectra of SN~1998bw. The cause of this shift is the growing opacity 
in the absorption bluewards of 4900\AA \ (rest wavelength). 

\begin{figure*}
  \includegraphics[height=.29\textheight]{030329V.eps}
  \includegraphics[height=.31\textheight]{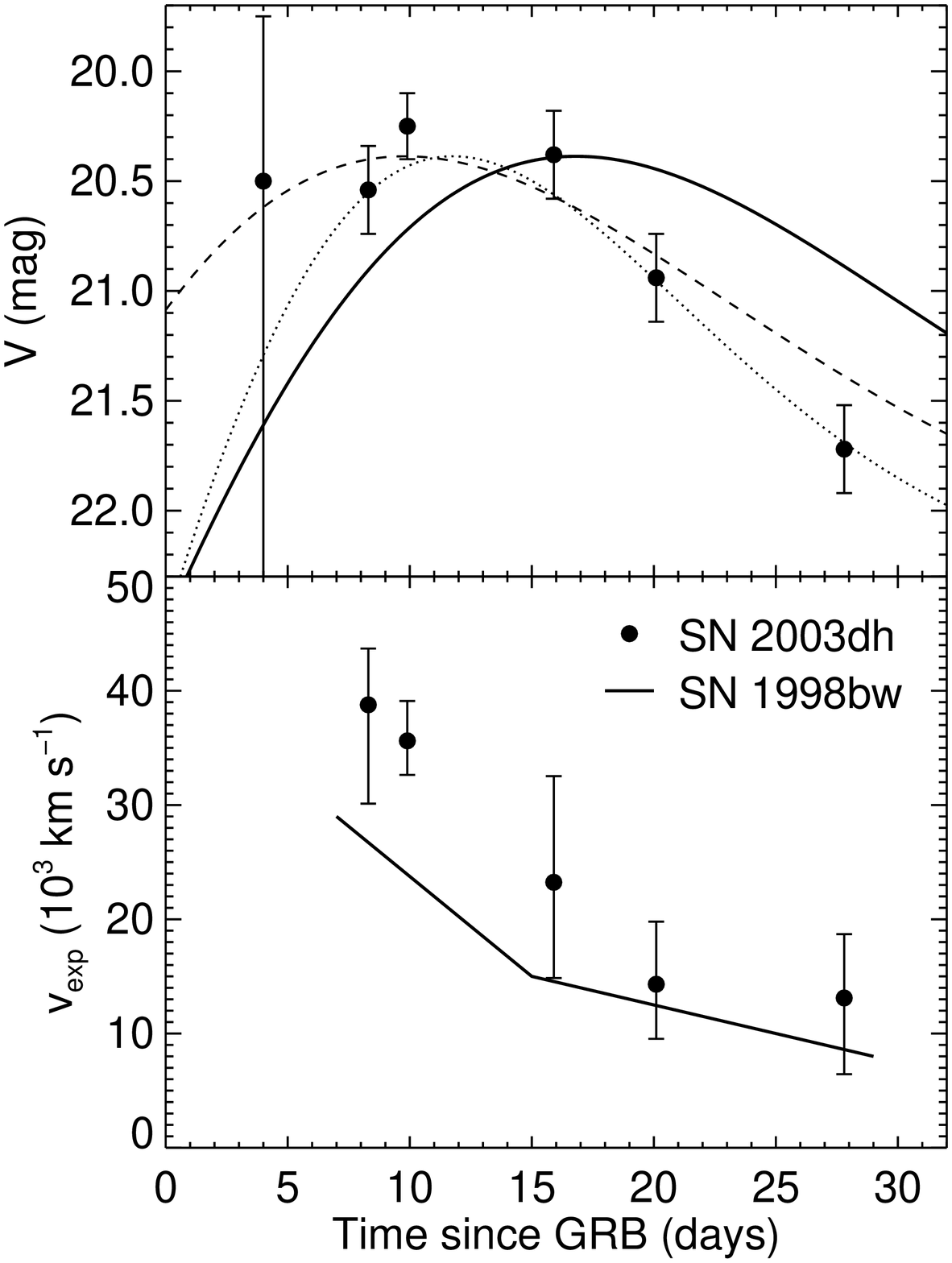}
  \caption{}
\end{figure*}

The spectral 
decompositions provide the fraction of the total flux in the V-band that is due 
to the supernova. The resulting SN~2003dh V magnitudes are plotted in the
right panel of Fig.~2. The full drawn line in this plot shows the brightness of 
SN~1998bw as it would have appeared in the V-band at z = 0.1685 as a function of 
time (restframe) since GRB~980425. Dashed line, as for the solid line but shifted 
7 days earlier. Such an evolution may be expected if the supernova exploded 
seven days before the GRB. For SN~2003dh, however, this is inconsistent with its 
spectral evolution (Fig. 2). Dotted line, as for solid line, but evolution 
speeded up by multiplying time by 0.7. A faster rise and decay may be expected 
in asymmetric models in which an oblate supernova is seen pole-on (e.g. Woosley
et al. in these proceedings). We assumed 0.20 mag extinction for SN~1998bw 
(Patat et al. 2001) and none for SN~2003dh. The bottom right plot in Fig.~2 shows 
the expansion velocities as a function of time in the restframe. Filled circles, 
SN~2003dh; solid line, SN~1998bw. The SN~2003dh values are our best estimates based 
on the decomposed spectra (see Hjorth et al. 2003 for details). We caution that 
in some cases these values are very uncertain owing to other features in the 
spectra around the expected minimum. The solid line shows the trend for SN~1998bw 
based on the data points in Patat et al. (2001). The consistent decaying trend 
in the ejecta velocity of SN~2003dh, together with its very high initial value, 
indicate that there was no delay between the GRB and the onset of the SN 
explosion.

Finally we note that the host galaxy is an actively star forming dwarf galaxy
with a total luminosity similar to that of the SMC and with a moderate 
metallicity ([O/H] $\approx-1$, Hjorth et al. 2003). From the OII and 
H$\alpha$ emission lines (shown along with the
other detected emission lines in the right panel of Fig.~1) we infer a star 
formation rate of a few tenths solar masses per year. In this respect 
GRB~030329/SN~2003dh is also similar to GRB~980425/SN~1998bw that was hosted 
by an actively star forming dwarf galaxy of type SBc (Fynbo et al. 2001).

\section{Comparison with independent studies}
A study of SN~2003dh has been presented in the paper by Stanek et al. (2003) 
and in the thorough and comprehensive paper by Matheson et al. (2003). These 
authors follow a decomposition strategy very similar to the one described 
above leading to very similar results for the properties of 
SN~2003dh. Two spectra during the supernova dominated phase were also 
secured by Kawabata et al. (2003). The most significant difference between 
our and other works is that Matheson et al. (2003) find a lightcurve for 
SN~2003dh that is identical to that of SN~1998bw within the erros. This is 
clearly not the case in the right panel of Fig.~2. Bloom et al.
(2003) speculate that there was a significant chromatic slit loss
in our April 3 observation due to the fact that we did not observe at the
parallactic angle. However, the FORS spectrographs are equipped with
atmospheric dispersion 
correctors\footnote{see http://www.eso.org/instruments/fors1/adc.html for 
details} that should minimize this effect. In a future paper in preparation
we will study in more detail if the lightcurve
of SN~2003dh is consistent with that of SN~1998bw or not.

\section{XRF~030723}
We end this contribution by describing our most recent results from
follow-up observations of the X-Ray Flash of July 23, 2003 (Fynbo 
et al. 2004, in preparation). XRF~030723 was located precisely by the 
SXC detector on Hete-II (Prigozhin et al. 2003). We started observations 
4 hrs after the burst and followed the evolution of the lightcurve during 
two months thereafter. In the left panel of Fig.~3 we show the R-band
lightcurve from a few hours to about 70 days after the explosion. 
The decay curve is consistent with being very flat during the first 24 hr
after the burst. Around 1 day after the burst the decay slope
$dm/d\log{t}$ steepens to about $-2$ and it remains so for the
following 4-5 days. Around a week after the burst the decay curve
starts to deviate from the fast decay and it then quickly
rises to a secondary maximum, peaked at around 16 days,
followed by a new steep decay. In the middle panel of Fig.~3 we
show the bump emission with an extrapolation of the afterglow
lightcurve subtracted.

Spectroscopic observations were secured on July 26 and on August 8 
during the steeply declining afterglow and bump phases respectively.
The spectrum of the afterglow from July 26 covers the region from about 
3800\AA \ to 8500\AA \ and it shows a featureless continuum with no 
significant absorption or emission lines. From the lack of Ly$\alpha$ absorption
we infer an upper limit to the redshift of about $z=2.1$. The spectrum
taken during the bump on August 8 covers the region 
5300\AA \ to 8600\AA. It also shows no significant narrow emission 
or absorption lines. 

Multicolor imaging was secured on four epochs. The afterglow spectral
energy distribution is well described by a $\beta \approx -1$ power-law
over the full range from the U-band to the K-band (right panel of
Fig.~3). During the bump phase the spectral energy distribution starts
to deviate strongly from a power-law shape due to a strong decrease in
the flux in the bluest bands.

In a sense the situation for XRF~030723 is opposite to that of GRB~030329.
For GRB~030329 the lightcurve did not show a strong bump apparently due
to a late break in the afterglow lightcurve that by coincidence balanced
out the extra emission from the SN. On the other hand the spectroscopic 
evidence was unambiguous. For XRF~030723 the photometry shows the most 
significant late time bump ever detected in an afterglow lightcurve at the 
time expected for an underlying supernova, but the spectroscopic evidence is 
more unclear. Nevertheless, it is clear that a SN~1998bw lightcurve is 
inconsistent with our data at any redshift. So far the best match found is 
for a SN similar to the type Ic SN~1994I, which had a very early peak time 
and a 
rather narrow peak in its lightcurve, at a redshift around $z\approx0.6$ 
(Fig.~3, middle panel). Interestingly, a SN similar to SN~1994I has also been 
proposed to be associated to GRB~021211 (Della Valle et al. 2003). However, even 
this SN has a too slow rise to match the data. 

\begin{figure*}
  \includegraphics[height=.2\textheight]{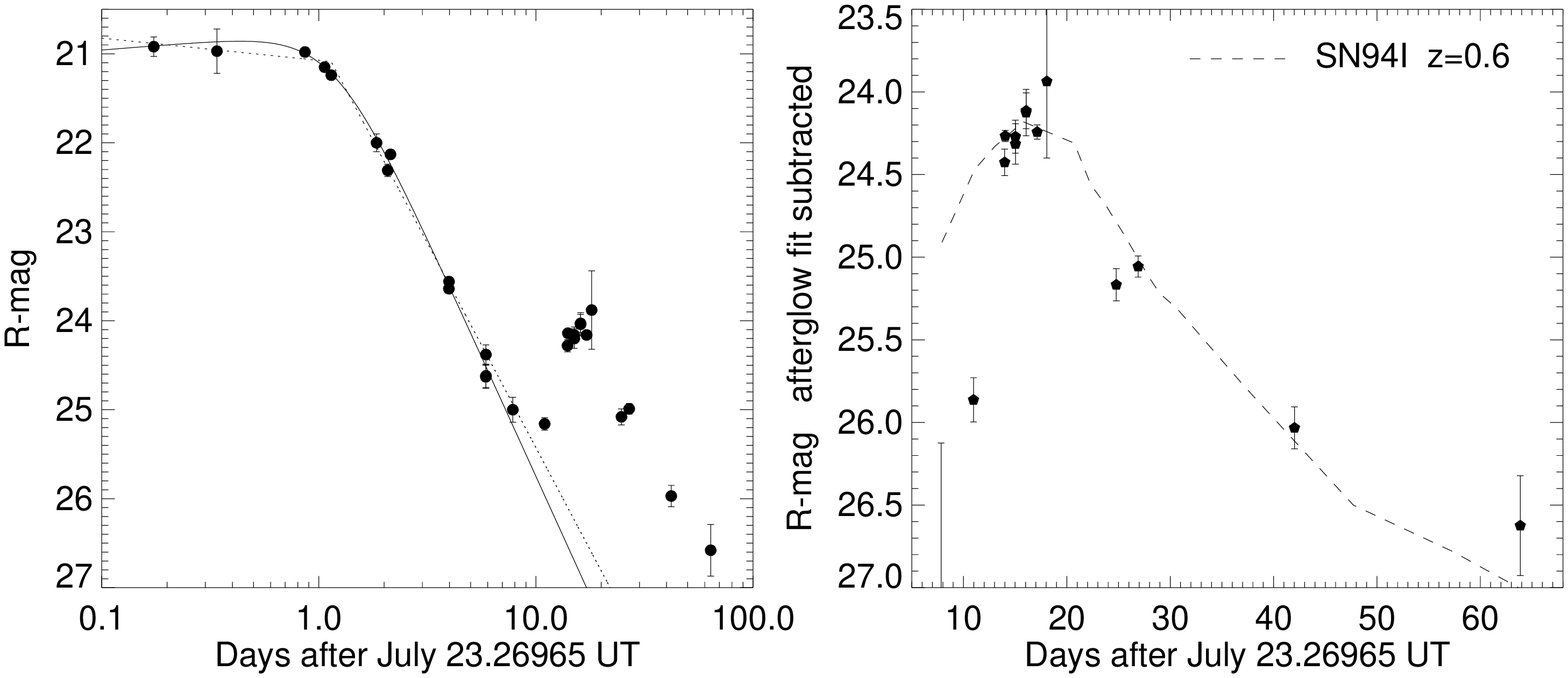} 
  \includegraphics[height=.2\textheight]{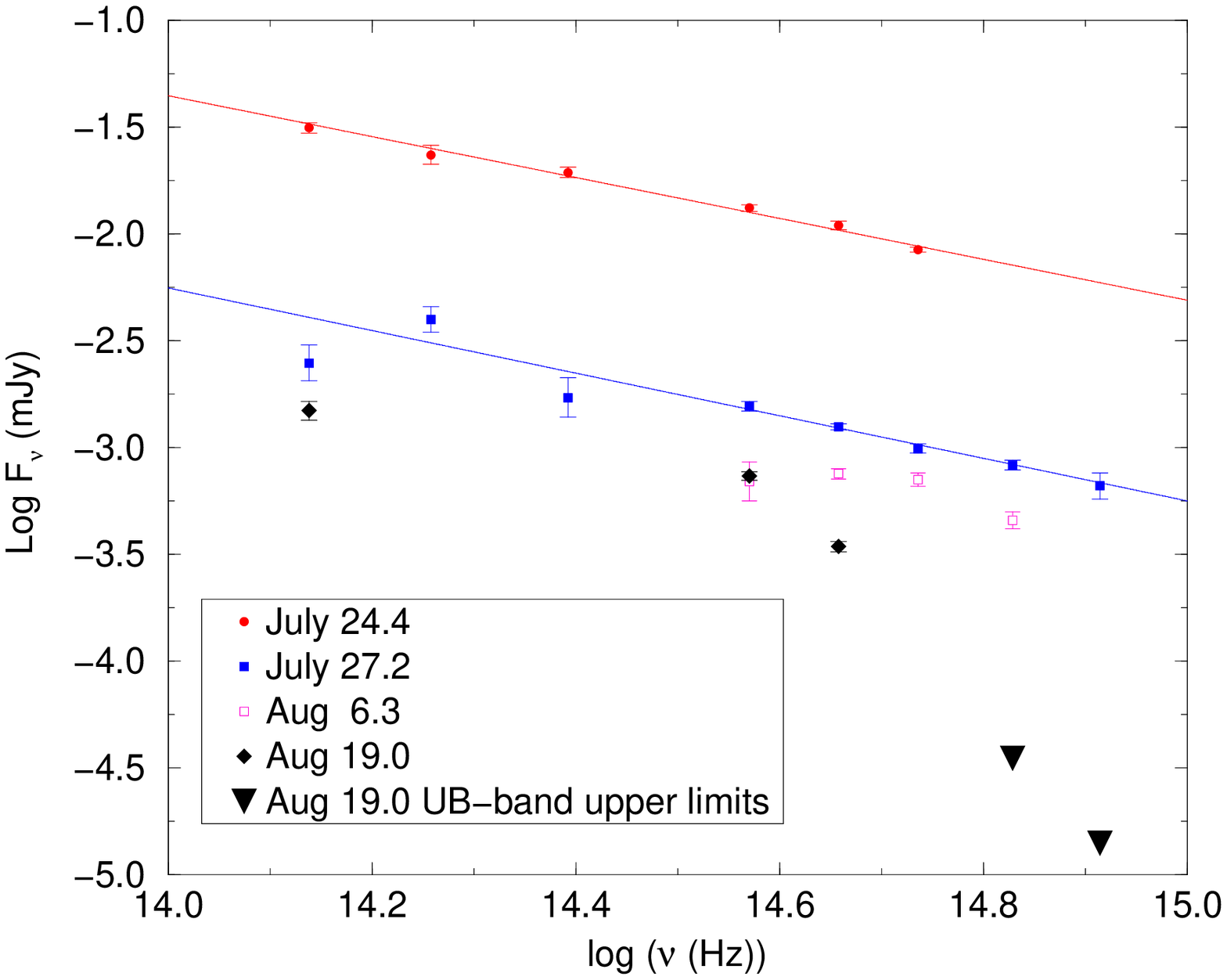}
  \caption{}
\end{figure*}

\section{Conclusion}
The connection between core collapse SN, more specifically of type Ib/c, and 
long duration GRBs has been demonstrated to everybody's satisfaction in the 
case of GRB~030329/SN~2003dh. The results for XRF~030723 show that XRFs are 
most likely also related to core collapse SN of type Ib/c. However, other 
potential explanations for the light curve bump should be investigated. A 
connection between XRFs and SN of type Ib/c would support the
hypothesis that XRFs and GRBs are manifestations of the same underlying 
phenomenon seen either under different viewing angles or with different 
baryon loadings.


\begin{theacknowledgments}
This paper is based on observations collected by the Gamma-Ray Burst 
Collaboration at ESO (GRACE) at the European Southern Observatory, Paranal, 
Chile. 
We acknowledge benefits from collaboration within the EU FP5 
Research Training Network "Gamma-Ray Bursts: An Enigma and a Tool". This work 
was also supported by the Danish Natural Science Research Council (SNF) and by
the Carlsberg Foundation.
\end{theacknowledgments}

\end{document}